\documentclass[a4paper,10pt]{revtex4-2}
\usepackage{mathrsfs}
\usepackage{graphicx}
\usepackage{latexsym}
\usepackage{amsmath}
\usepackage{amssymb}
\usepackage{textcomp}
\usepackage{amsbsy}
\usepackage{graphics}
\usepackage{epstopdf}
\usepackage{color}

\allowdisplaybreaks[4]

\begin{document}

\tolerance=5000

\title{Little Rip and Pseudo Rip cosmological models with coupled dark energy based on a new generalized entropy    }

\author{I.~Brevik,$^{1}$\,\thanks{iver.h.brevik@ntnu.no}
A.~V.~Timoshkin,$^{2,3}$\,\thanks{alex.timosh@rambler.ru}
}
 \affiliation{ $^{1)}$ Department of Energy and Process Engineering,
Norwegian University of Science and Technology, N-7491 Trondheim, Norway. ORCID: 0000-0002-9793-8278\\
$^{2)}$Institute of Scientific Research and Development, Tomsk State Pedagogical University (TSPU),  634061 Tomsk, Russia. ORCID: 0000-0002-5208-8687 \\
$^{3)}$ Lab. for Theor. Cosmology, International Centre  of Gravity and Cosmos,  Tomsk State University of Control Systems and Radio Electronics
(TUSUR),   634050 Tomsk, Russia
}

\tolerance=5000

\begin{abstract}
We study Little Rip (LR) and Pseudo Rip (PR) cosmological models containing two coupled fluids: dark energy and dark matter. We assume a spatially flat Friedmann-Robertson-Walker (FRW) universe. The interaction between the dark energy and the dark matter fluid components is described in terms of the parameters in the generalized equation of state (EoS) in presence of the bulk viscosity. We consider entropic cosmology and use a description based on a new generalized entropy function, which was proposed by Nojiri-Odintsov-Faraoni [1]. Conditions for the appearance of the (LR) and the (PR) in terms of the parameters of the (EoS) are obtained. Introducing an energy density $\rho_g$ corresponding to a specified entropy function $S_g$, together with an interaction term $Q$ in the gravitational equations of motion, we derive modified forms of the EoS parameters. We discuss the corrections of the thermodynamic parameters associated with the generalized entropy function. Properties of the late universe as well as in the early universe  in this formalism are pointed out.

\bigskip



\end{abstract}

\date{\today}
\maketitle

\section{Introduction}

The appearance of new theoretical models for dark energy is associated with the discovery in 1998, using astronomical observational data, obtained independently in the laboratories of A. Riess and S. Perlmutter, of the accelerated expansion of the modern universe \cite{2,3}.  For a universe filled with phantom energy, when the thermodynamic parameter  (EoS) lies in the regime  $(\omega < -1)$,  there are many possible  scenarios for the end of such a universe. Phantom dark energy can lead to a Big Rip (BR) future singularity, when the scale factor is finite at the rip time (for a general classification of finite-time future singularities see \cite{4,5,6,7,triveditimoshkin23,timoshkinyurov23,breviktimoshkin23}).

 However, a final evolution without singularities is also possible. If the parameter $\omega$  asymptotically tends to $-1$  and the energy density increases with time or remains constant, no finite-time future singularity will be ever formed \cite{6,7,8,9}.
If the cosmic energy density remains constant or increases monotonically in the future, then all the possible fates of our universe can be divided into four categories based on the time asymptotic regimes of the Hubble function \cite{7}:

\noindent 1. Big Rip: $H \rightarrow \infty$ when $t\rightarrow t_{\rm rip} <\infty$;

\noindent 2. Little Rip: $H \rightarrow \infty$ when $t\rightarrow \infty$;

\noindent 3. Hubble parameter  $H=$constant;

\noindent 4.  Pseudo-Rip: $H(t) \rightarrow H_\infty $ when $t\rightarrow \infty$, with $H_\infty$ a constant.

\bigskip
The theory predicts many interesting scenarios for the evolution of the universe in the future, including the (BR) \cite{10}, the (LR) \cite{11,12}, the (PR) \cite{13} and the Quasi Rip \cite{14} cosmological models.  Moreover, in the early universe there is the possibility of realizing not only inflationary acceleration, but also bounce model for the matter \cite{15}.

The (BR) singularity means that the basic physical quantities become infinity during a finite time of evolution.   In the (LR) scenario, it takes infinite time to reach the singularity. In the (PR) cosmology, the Hubble parameter tends to a “cosmological constant” in the far future. In phenomena with a discontinuity, like the (LR) and the (PR), the thermodynamic parameter  $\omega$ asymptotically tends to the value $-1$. These models are based on the assumption that the dark energy density is a monotonically increasing function. The evolution of the universe for bounce cosmology is associated with the absence of singularities. In the cosmological model with matter bounce, the universe passes from an era of accelerated compression to an era of expansion through a bounce without encountering   a singularity, what implies a model of a cyclic universe \cite{16}.

An important topic in modern cosmology is  the interaction between dark energy and dark matter in relation to the accelerated cosmic expansion. This can be studied by including the coupling of dark energy to dark matter. One of the most successful methods for such a description  is the generalized (EoS) \cite{17,18,19,20}.

An alternative approach to the modified theory of gravity is based on the connection between gravity and thermodynamics \cite{1,21}.  According to entropic cosmology, the entropy  generates the energy density and the pressure in the Friedmann equations. As a result, Friedmann's equations are a consequence of the fundamental laws of thermodynamics. In addition, entropy cosmology provides a good description of the inflationary phase with a smooth exit, with theoretical predictions of observed indices consistent with recent Planck data for relevant entropy energy ranges \cite{22,23,24}.  The nonsingular property of the entropy function plays also an important role in describing bounce cosmology. From the thermodynamic approach it follows, that at the late stage of the evolution of the universe, assuming  the cosmological constant to be  zero, the dark energy density is determined  entirely by  the entropy energy density. In this case the theoretical values of the parameters in the (EoS) are compatible with the data of astronomical observations from the Planck satellite. If one includes  a cosmological constant in the entropy model, this makes the theory  more viable. In this   generalized entropy cosmology,   the cosmological constant plays the role of  a special case.

The purpose of the present paper is to obtain analytical representations in cosmological models with (LR) and (PR), induced by an inhomogeneous viscous fluid, using the thermodynamic parameter and bulk viscosity in the generalized equation of state.  We will study an  entropy cosmology model in a spatially flat Friedmann-Robertson-Walker metric.

\section{Generalized entropy}

In Refs.~\cite{21,5} the generalized entropy $S_g$ was given as
\begin{equation}
S_g(\alpha_+, \alpha_-, \beta, \gamma)=
\frac{1}{\gamma}\left[ \left(1+\frac{\alpha_+}{\beta}S\right)^\beta -\left(1+\frac{\alpha_-}{\beta}S\right)^{-\beta}\right], \label{1}
\end{equation}
where $\alpha_+, \alpha_-, \beta,\gamma$ are positive parameters and $S$ is the Bekenstein-Hawking entropy. The latter is the thermodynamic entropy of a black hole \cite{25,26},
\begin{equation}
S= \frac{A}{4G}, \label{2}
\end{equation}
where $A = 4\pi r_h^2$ is the horizon area and $r_h$ the horizon radius. The Bekenstein-Hawking entropy describes the radiation from a black hole.

Recently, various forms of entropy other the the Bekenstein-Hawking entropy have been proposed, such as the Tsallis \cite{27} and the R{\'e}nyi \cite{28} entropies. There  are proposals also by Barrow \cite{29} and by Kaniadakis \cite{kaniadakis05}. These entropies have common properties: in limiting cases they reduce to the Bekenstein-Hawking entropy and are monotonically increasing functions of the Bekenstein-Hawking entropy variable. It has been proven \cite{5,21} that the generalized entropy function $S_g$ reduces to all the entropies indicated above, for suitable choices of the parameters. This implies that the minimum number of parameters in the generalized entropy function, able to include all kown entropies, is four \cite{5,21}.

\section{ Viscous entropic cosmology coupled with dark matter}

We consider the flat FRW universe filled with two interacting fluids: dark energy, and dark matter,
\begin{equation}
ds^2 = -dt^2 +a^2(t)\sum_{i=1,2,3}(dx^i)^2, \label{3}
\end{equation}
where $a(t)$ is the scale factor. The entropy function is taken to be the four-parameter form $S_g$ given in Eq.~(\ref{1}).

The modified Friedmann equation  becomes
\begin{equation}
H^2 = \frac{k^2}{3}(\rho + \rho_m +\rho_g), \label{4}
\end{equation}
where $H= \dot{a}(t)/a(t)$ is the Hubble function, $k^2= 8\pi G$ is Einstein's gravitational constant with $G$ the Newton gravitational constant;  $p,\rho$ and $p_m, \rho_m$ are the pressure and the energy density of dark energy and dark matter, respectively. A dot means derivative with respect to the cosmic time $t$. Moreover, $\rho_g$ nd $p_g$ mean the energy density and the pressure corresponding to the entropy function $S_g$.

Next, we consider cosmological implications of the modified Friedmann equation. We write the dynamic equations in the form \cite{4}
\newpage
\begin{equation*}
\dot{\rho}+3H(p+\rho) = -Q,
\end{equation*}
\begin{equation}
\dot{\rho}_m + 3H(p_m + \rho_m)= Q, \label{5}
\end{equation}
\begin{equation*}
\dot{H}= -\frac{k^2}{2} (p+\rho + p_m + \rho_m),
\end{equation*}
where $Q$ is the interaction term between dark energy and dark matter.

Let us now apply the generalized entropy function $S_g$ to the (LR) and the (PR) models. In the late universe, the parameter $\gamma$ becomes constant, $\gamma =\gamma_0$. As a result, the entropy function takes the following form, \cite{21}
\begin{equation}
S_g = \frac{1}{\gamma_0}\left[ \left(1+\frac{\alpha_+}{\beta}S\right)^\beta -\left(1+\frac{\alpha_-}{\beta}S\right)^{-\beta}\right], \label{6}
\end{equation}
with $S= \pi/(GH)^2$.

At a late stage in the evolution of the universe, the estimate of the Hubble function is of order $10^{-40}~$GeV and the condition $GH^2 \ll 1$ is satisfied. That is why the expression for the energy density $\rho_g$ corresponding to the entropy function (\ref{6}) can be approximated by the following formula,  \cite{21}
\begin{equation}
\rho_g = \frac{3H^2}{k^2}\left[ 1-\frac{\alpha_+}{\gamma_0(2-\beta)}\left( \frac{GH^2\beta}{\pi \alpha_+}\right)^{1-\beta} \right]. \label{7}
\end{equation}
We will now investigate the (LR) and the (PR) cosmological models.

\section{Little Rip and Pseudo Rip cosmologies from generalized entropy function}

\subsection{Little Rip model: first example}

A characteristic feature of (LR) cosmology is that the dark energy density increases asymptotically with time, so it takes an infinite time to reach a singularity. In this case the thermodynamic parameter (EoS) is less than minus one $(\omega <-1)$, but tends to minus one asymptotically.  This is thus a soft version of the future singularity.

Let us consider a (LR) model with the following form for the Hubble function, \cite{7}
\begin{equation}
H(t)= H_0 e^{\lambda t}, \quad H_0>0,~~\lambda >0. \label{8}
\end{equation}
We will consider the evolution of the universe from the instant $t=0$ onwards, where this instant lies in the epoch of the very early universe. The symbol $H_0$ denotes the Huble function at present time.

We will in the following assume that the dark matter is dust matter, so that $p_m=0$ and the gravitational equation of motion for dark matter reduces to
\begin{equation}
\dot{\rho}_m+3H\rho_m=Q. \label{9}
\end{equation}
We take the same coupling as in Ref.~\cite{brevik15},
\begin{equation}
Q= \delta H \rho_m, \label{10}
\end{equation}
where $\delta$ is positive nondimensional constant. The solution of Eq.~(\ref{9}) for dark matter is
\begin{equation}
\rho_m(t)= \rho_m^{(0)} \exp\left( \frac{\delta -3}{\lambda}H\right), \label{11}
\end{equation}
where $\rho_m^{(0)}$ is a integration constant. If $\delta$ is small $(<3)$, then $\rho_m\rightarrow 0$ when $t\rightarrow +\infty$.

Let us consider the simple case where $\beta=1$  in the expression  (\ref{7})  for the energy density $\rho_g$. We then obtain
\begin{equation}
\rho_g= \frac{3H^2}{k^2}\left( 1-\frac{\alpha_+}{\gamma_0}\right). \label{12}
\end{equation}

Now consider the following representation of the (EoS) of the viscous fluid \cite{19},
\begin{equation}
p= \omega (\rho,t)\rho -3H\zeta (H,t), \label{13}
\end{equation}
where $\omega(\rho,t)$ is the thermodynamic parameter and $\zeta(H,t)$ the bulk viscosity. From thermodynamic considerations, it follows that $\zeta(H,t)>0$.

Taking into account Eqs.~(\ref{4},\ref{8},\ref{10},\ref{12},\ref{13}) we obtain the gravitational equation of motion for dark energy,
\begin{equation}
\frac{2\alpha_+\lambda}{\gamma_0k^2}H - \left(\frac{\delta}{3} -1\right)\rho_m +[\omega(t)+1]\left( \frac{3\alpha_+}{\gamma_0k^2}H^2-\rho_m\right)
-3H\zeta(H,t)= \frac{\delta}{3}\rho_m. \label{14}
\end{equation}
From this equation we derive the following expression for the thermodynamic parameter,
\begin{equation}
\omega(t)= -1-\frac{\left(\frac{2\alpha_+\lambda}{\gamma_0k^2} -3\zeta(H,t)\right) + \left( 1-\frac{2}{3}\delta\right)\rho_m}
{ \frac{3\alpha_+}{\gamma_0k^2}H^2 - \rho_m}. \label{15}
\end{equation}
Here the first term in the numerator is a constant; the second term contains the contribution from the coupling and the entropy function. The (L,R) behavior in this case is caused by the function $\zeta(H,t)$.

Let us consider the approximation $\rho_m=0$ (the influence from dark matter not taken into account). Then,
\begin{equation}
\omega(t)= -1-\left[ \frac{2}{3}\lambda -\frac{\gamma_0k^2}{\alpha_+}\zeta(H,t)\right]H^{-1}. \label{16}
\end{equation}
First, we put here the bulk viscosity equal to a constant, $\zeta(H,t)=\zeta_0>0$. Then in the limit $t\rightarrow \infty, H\rightarrow \infty$ and $\omega(t)\rightarrow -1$ (the case of a cosmological constant).

Second, we put the viscosity proportional to the Hubble function, $\zeta(H,t)= 3\tau H$, where $\tau$ is a positive constant. We the obtain
\begin{equation}
\omega(t)=-1-\frac{2\lambda}{H}+\frac{3\tau \gamma_0k^2}{\alpha_+}. \label{17}
\end{equation}
Thus in the limit $t\rightarrow \infty, \omega(t) \rightarrow -1+\frac{3\tau \gamma_0k^2
}{\alpha_+} >-1$ (the universe is in the nonphantom phase). It follows that both the entropy density and the bulk viscosity influence the transitions between different phases of the phantom expansion.

Solving for the bulk viscosity $\zeta(H,t)$ from Eq.~(\ref{14}), we obtain
\begin{equation}
\zeta(H,t)=\frac{2}{3}\frac{\lambda \alpha_+}{\gamma_0k^2} +\frac{1}{3H}\left\{ [\omega(t)+1]\left( \frac{3\alpha_+}{\gamma_0k^2}H^2-\rho_m\right)
+ \left(1-\frac{2}{3}\delta\right) \rho_m\right\}. \label{18}
\end{equation}
It follows that the entropic energy density (first term) increases the viscosity.

In order to see better the influence from the entropic energy density on the viscosity, we return again to the approximation where dark matter is omitted, in which case
\begin{equation}
\zeta(H,t)=\frac{\alpha_+}{\gamma_0k^2}\left\{ \frac{2}{3}\lambda +[\omega(t)+1]H \right\}. \label{19}
\end{equation}
When $t\rightarrow \infty, H \rightarrow \infty$ and the bulk viscosity increases.

Thus, we have formulated the (L,R) theory of cosmology taking into account the bulk viscosity, as well as the interactions between dark energy and dark matter, in the entropic cosmological model.

\subsection{Little Rip model: second example}

We now consider another (LR) model where the Hubble function takes the form \cite{7}
\begin{equation}
H(t)= H_0\exp{(Ce^{\lambda t})}, \label{20}
\end{equation}
where $H_0,C$ and $\lambda$ are positive constants.

Note that in this case the Hubble function increases monotonically faster than in Eq.~(\ref{8}). We take the interaction term $Q$ in the same form (\ref{10}) as before. Soving the the gravitational equation of motion (\ref{5}) for dark matter, we find
\begin{equation}
\rho_m(t)= \rho_m^{(0)} \exp{\left[ \frac{\delta-3}{\lambda}H_0 Ei(Ce^{\lambda t}) \right]}, \label{21}
\end{equation}
In the limit $t\rightarrow 0$ (early universe), $ \rho_m(t)\rightarrow  \rho_m^{(0)} \exp{\left[ \frac{\delta-3}{\lambda}H_0 Ei(C) \right]}. $ If $\delta$ is small $(<3)$ and the parameter tends to $C\rightarrow 0+$, then the dark energy $\rho_m \rightarrow 0$. This implies a physical interpretation of the moment $t=0$ in this model: this is the time when dark matter begins to appear in the universe. In the opposite case, when $t\rightarrow +\infty$  (late universe), the dark energy density increases, $\rho_m \rightarrow +\infty$.

Now taking into account Eqs.~(\ref{4},\ref{10},\ref{12},\ref{13},\ref{20}) we can write the equation of  motion for dark energy as
\begin{equation}
2\frac{\alpha_+\lambda}{\gamma_0k^2}Ce^{\lambda t}+\omega(t)\left(\frac{3\alpha_+}{\gamma_0k^2}H-\frac{\rho_m}{3H}\right) +\frac{\alpha_+}{\gamma_0k^2}H -\zeta(H,t)=0. \label{22}
\end{equation}
From this we obtain for the bulk viscosity
\begin{equation}
\zeta(H,t)= \frac{\alpha_+}{\gamma_0k^2}\left\{ \frac{2}{3}C\lambda e^{\lambda t}+[\omega(t)+1]H\right\} -\omega(t)\frac{\rho_m}{3H}. \label{23}
\end{equation}
The first term to the right in this expression contains the contribution from the entropy function; the second describes the contribution from the coupling with dark matter. Thus, we have in this way formulated the (L,R) cosmology theory when including the bulk viscosity $\zeta(H,t)$.

Let us solve Eq.~(\ref{22}) with respect to the thermodynamic parameter $\omega(t)$:
\begin{equation}
\omega(t)= \frac{ \frac{\zeta(H,t)}{H}-\frac{\alpha_+}{\gamma_0k^2}\left( \frac{2}{3}\lambda C\frac{e^{\lambda t}}{H}+1\right) }
{\frac{\alpha_+}{\gamma_0k^2}-\frac{\rho_m}{H^2} }. \label{24}
\end{equation}
If we assume that $\zeta(H,t) = \zeta_0$ is a constant, then at $t\rightarrow \infty, H\rightarrow \infty$ and $\omega \rightarrow -1$. In this case the dark energy has the structure of a cosmological constant, meaning that our universe tends to the de Sitter universe.

In the other case discussed above, when the bulk viscosity was proportional to the Hubble function, $\zeta(H,t)=3\tau H$ with $\tau$ a positive dimensional constant, we obtain in the limit $t\rightarrow \infty$
\begin{equation}
\omega(t) \rightarrow -1+\frac{3\tau \gamma_0k^2}{\alpha_+} > -1. \label{25}
\end{equation}
The universe is in a nonphantom phase.

\subsection{Pseudo Rip model: first example}

Let us study models where the dark energy density increases monotonically. In this case, the expansion of the universe approaches the exponential regime asymptotically. The Hubble function tends to a constant value (cosmological constant), that means, a de Sitter space.

We will analyze this model in analogy with the above (LR) model. Let us suppose that the Hubble function has the form \cite{13}
\begin{equation}
H= H_0-H_1e^{-\lambda t}, \label{26}
\end{equation}
where $H_0, H_1$ and $\lambda$ are positive constants. We assume that $H_0>H_1$ when $t>0$.

We suppose that the interaction between dark energy and dark matter has the form \cite{brevik15}
\begin{equation}
Q= \frac{3Q_0} {H_0}He^{\lambda t}, \label{27}
\end{equation}
with $Q_0$ an unspecified constant.

We now take $\lambda =H_0$, and solve the gravitational equation of motion (\ref{5}) for dark matter with the result
\begin{equation}
\rho_m(t)= \frac{Q_0}{H_0}\left\{ e^{az}\sum_{k=0}^2 \frac{a^k}{(3-k)!z^{(3-k)}}\left[ 3\sum_{n=0}^2 \frac{a^{n+3}}{(3-n)!z^{(3-n)}} -a\right]
-\frac{a^4}{3!}Ei(az)\left( \frac{a^2}{60}-1\right)\right\} z^3e^{az}, \label{28}
\end{equation}
where $a= 3H_1/H_0$ and $z=e^{-H_0t}$.

If $t\rightarrow \infty, z\rightarrow 0$ and $\rho_m \rightarrow \infty$. It means that dark matter increases in the late universe. In the opposite case when $t\rightarrow 0, z\rightarrow 1$, and the dark matter energy density tends to a constant value
\begin{equation}
\rho_m(0)= \frac{Q_0}{H_0}e^a \left\{  \sum_{k=0}^2 \frac{a^k}{(3-k)!}\left[ 3\sum_{n=0}^2 \frac{a^{n+3}}{(3-n)!} -a\right]
-\frac{a^4}{3!}Ei(a)\left( \frac{a^2}{60}-1\right)\right\}. \label{29}
\end{equation}
Taking into account Eqs.~(\ref{4},\ref{12},\ref{13},\ref{26},\ref{27}), the dark energy conservation law becomes
\begin{equation}
\frac{\alpha_+}{\gamma_0}H_0(H_0-H)-\frac{\dot{\rho}_m}{3H}
 +\omega(t)\left( \frac{3\alpha_+}{\gamma_0k^2}H^2-\rho_m\right) -3H\zeta(H,t)+\frac{Q_0}{H_0z}=0. \label{30}
\end{equation}
The expression for the thermodynamic parameter,
\begin{equation}
\omega(t)= -\frac{ \frac{\alpha_+}{\gamma_0}H_0H_1z -\frac{\dot{\rho}_m}{3H}+3H\zeta(H,t)+\frac{Q_0}{H_0z} }
{ \frac{3\alpha_+}{\gamma_0k^2}H^2-\rho_m}, \label{31}
\end{equation}
shows that the Pseudo-Rip behavior is determined by the viscosity and entropy functions.

Thus we see that, unlike the case where only a dark energy was involved, the interaction between dark energy and dark matter through the generalized entropy function leads to corrections for the thermodynamic parameter.

\subsection{Pseudo Rip model: second example}

Next, we consider a cosmological model, asymptotically describing the de Sitter evolution \cite{34}, where the Hubble function is \cite{6}
\begin{equation}
H(t)= \frac{x_f}{\sqrt{ 3} }\left[ 1-\left( \frac{x_0}{x_f}\right) \exp{\left( -\frac{\sqrt{3}At}{2x_f}  \right)} \right], \label{32}
\end{equation}
where $x_0=\sqrt{\rho_0}$ is the energy density at present, $x_f$ is a finite value related to divergence in the cosmic time, and $A$ is a positive dimensional constant. When $t\rightarrow \infty$, the Hubble function $H\rightarrow x_f/\sqrt{3}$ (cosmological constant), and the expression (\ref{32}) asymptotically tends to the de Sitter solution.

Let us take the interaction term in the form
\begin{equation}
Q=Q_0\exp{\left[  \frac{2x_f^2}{A}\exp{\left(  -\frac{\sqrt{3}At}{2x_f}   \right) }      \right] }, \label{33}
\end{equation}
where $Q_0$ is a constant.

Let us put $A= x_f^2$. Then the solution of the gravitational equation (\ref{5}) becomes
\begin{equation}
  \rho_m(t)= \exp{\left[ -\left( 1-\frac{x_0}{x_f}\right)\tau\right] }\left\{ \rho_0\tau +\frac{Q_0}{\sqrt{3}\,x_f}\left[ e^\tau -\tau Ei(\tau)\right] \right\}, \label{34}
\end{equation}
where $\tau = e^{-\sqrt{3}x_ft}$.

If $t\rightarrow \infty$, then $t\rightarrow 0$ and $\rho_m \rightarrow \frac{Q_0}{\sqrt{3}\,x_f}.$ The dark matter energy density tends to a constant value in the late universe. In the early universe, when $t\rightarrow 0, \tau\rightarrow 1$, we obtain
\begin{equation}
\rho_m(0)= \exp{\left[ -\left( 1-\frac{x_0}{x_f}\right)\right] }\left\{ \rho_0\tau +\frac{Q_0}{\sqrt{3}\,x_f}\left[ e - Ei(1)\right] \right\}. \label{35}
\end{equation}
Next, we write the energy conservation law, based on Eqs.~(\ref{4},\ref{12},\ref{13},\ref{32},\ref{33}), in the form
\begin{equation}
\frac{6\alpha_+}{\gamma_0k^2}x_f(x_0-x_f)\tau H-  \exp{\left[ -\left( 1-\frac{x_0}{x_f}\right)\right] }\tilde{\rho}_m+3H\left[[\omega(t)+1]\left(\frac{3\alpha_+}{\gamma_0k^2}H^2-\rho_m\right)
-3H\zeta(H,t)\right]=Q_0e^\tau, \label{36}
\end{equation}
where $ \tilde{\rho}_m= (x_f-x_0)\tau \left\{ \sqrt{3}\,\rho_0\tau +\frac{Q_0}{x_f}[\tau-Ei(\tau)]\right\} +\tau[ \sqrt{3}\,\rho_0\tau x_f-Q_0Ei(\tau)]. $

Let us consider the asymptotic case $t\rightarrow \infty$ (the late universe). Then $\tilde{\rho}_m \rightarrow 0$ and the equation (\ref{36}) simplifies to
\begin{equation}
[\omega(t)+1]\left( \frac{\alpha_+x_f^2}{\gamma_0k^2}-\frac{Q_0}{\sqrt{3}\,x_f}\right) -\sqrt{3}\, x_f\zeta(H,t)=\frac{Q_0}{\sqrt{3}\, x_f}. \label{37}
\end{equation}
Hence, we obtain the following expression for the thermodynamic parameter,
\begin{equation}
\omega(t)= -1- \frac{1+\frac{3x_f^2}{Q_0}\zeta(H,t)}{1-\frac{\sqrt{3}\,\alpha_+x_f^3}{\gamma_0k^2Q_0}}. \label{38}
\end{equation}
This is thus the result from the (PR) model.

Repeating the procedure again, we obtain the following expression for the bulk viscosity in this model,
\begin{equation}
\zeta(H, t)= \frac{1}{\sqrt{3}\, x_f}\left\{ [\omega(t)+1]\frac{\alpha_+ x_f^2}{\gamma_0k^2}-\frac{Q_0}{\sqrt{3}\, x_f} \right\}. \label{39}
\end{equation}
This expression contains a contribution from the generalized entropy function, as before.

	\section{ Conclusion}

In the present paper we have studied examples of (LR) and (PR) cosmological models described in flat (FRW) space time, when the cosmic fluid is viscous and is coupled with dark matter. We have found corrections in the thermodynamic parameter, due to the bulk viscosity, in the (EoS) for the dark energy. Previously, similar issues were discussed in \cite{brevik13,brevik15}.  The novelty of this work lies in the application of a thermodynamic approach based on the generalized entropy function. Including the energy density corresponding to the generalized entropy function in the Friedmann equation, leads us to modify  the parameters in the (EoS). For example, a consequence of the  corrections in the  cosmology of the (LR) or (PR) types through the thermodynamic parameter,  leads to a change in the expansion phase of the universe.

	It is of interest to note   that the generalized (EoS) cosmology may be  rewritten as  modified gravity \cite{nojiri06}. There is much interest currently to modified gravity; see reviews \cite{nojiri11,nojiri17}.  	It would be of interest to develop cosmological models of (LR) and (PR) cosmologies in modified gravity. This will be considered elsewhere.

	Finally, we mention that  the  relationship between   black holes and  generalized entropy has recently been discussed in Refs.~\cite{41,42}.

\end{document}